# Static Universe: Infinite, Eternal and Self-Sustainable

E. López-Sandoval
Centro Brasileiro de Pesquisas Físicas,
Rua Dr. Xavier Sigaud, 150
CEP 22290-180, Rio de Janeiro, RJ, Brazil.
elsandoval202020@gmail.com

*The task is not to see what has not been seen before, but to think what has never been thought before about what you see every day.*
*Erwin Schrödinger*

In this paper, we present a "stellar dynamics" model of an infinite Universe, where matter distribution follows an inverse proportionality squared relationship with respect to the distance from the rotation center of galaxy clusters and superclusters (which share a common rotation center). We assume the Universe has infinite similar centers in terms of structure and dynamic equilibrium. We consider stars in galaxies to be homogeneously distributed with spherical symmetry and average radius, and the same applies to galaxies in the Universe. We study the smoothed potential of this universe and examine the effect of gravity on starlight: by applying the equivalence principle, we derive a mathematical expression for Hubble's law and a formula for its redshift, potentially explaining this phenomenon as a gravitational effect. We also provide an approximate calculation of Cosmic Background Radiation (CBR), assuming this radiation is the light from all the universe's stars reaching us with an extreme redshift caused by gravity.

## 1. INTRODUCTION

"How is the Universe structured? Is it eternal, or did it have a beginning? Will it end, or is it infinite? These are questions that have puzzled humans, given our consciousness and position within it. Throughout history, we've developed various cosmogonies in an attempt to answer these questions.

The first scientific theory to provide plausible answers and form the basis of a cosmology was Newtonian Mechanics. Newton proposed a universe with Euclidean geometry and absolute time, where physical phenomena occur in a passive space, governed by his laws of motion and universal gravitation. In this model, massive bodies interact through instantaneous gravity, with no intermediary force.

Newton also theorized that the Universe was static and infinite, with matter in a state of balance due to its uniform and infinite distribution. Each star, in this model, is balanced by its neighbors and the universe as a whole, creating a locally unstable equilibrium, akin to vertically placed needles."

Regarding the quote from Newton, in a correspondence maintained with Richard Bentley [1]:

*The Lord affirms that all particles of matter in an infinite space have an infinite amount of matter of all sides and, consequently, an infinite attraction by all parts, having therefore to remain in balance because all the infinities are equal.*

In a subsequent letter, Newton concurred with Bentley's concept of this infinite system being in balance yet unstable, comparing it to vertically aligned needles. This analogy highlights the delicacy of the equilibrium, where minimal disturbance might lead to instability:

*Therefore, when I say that the equally dispersed matter by all the space would be added by its gravity in one or but immense masses, I would understand that this would be a matter that would not remain in rest in a precise balance.*

Newton's mechanical and fragmented perspective of the Universe persisted for roughly two centuries until Einstein (1917) introduced his gravitational theory in the General Theory of Relativity. In the Einstein's vision of the Universe, space, time, and matter are interconnected and continuous, influencing one another in a global, dynamic evolution [2]:

*When forced to summarize the general theory of relativity in one sentence:* ***Time*** *and* ***space*** *and* ***gravitation*** *have no separate existence from* ***matter****. ... Physical objects are not in space, but these objects are spatially extended. In this way the concept 'empty space' loses its meaning. ... Since the theory of general relativity implies the representation of physical reality by a continuous field, the concept of particles or material points cannot play a fundamental part, ... and can only appear as a limited region in space where the field strength / energy density are particularly high.*

Einstein's theory introduced a four-dimensional Riemannian geometry encompassing three spatial dimensions and one temporal dimension. He formulated a field equation that correlates matter, energy distribution, and space-time curvature. Furthermore, he introduced the Perfect Cosmological Principle, positing that matter distribution in the Universe is irregular on a small scale but achieves homogeneity at large scales.

By applying his equation to the Universe as a whole, Einstein described its evolution, simplifying the complexity of the equations and enabling solutions for otherwise intractable problems. He also introduced a cosmological constant, representing a repulsive force counteracting gravity and preventing all matter from collapsing towards a central point. This allowed for the description of a static Universe, free from global motion or expansion relative to a balance center.

Later, in 1922, Friedmann discovered that the static Universe wasn't the only solution, revealing that the cosmological constant couldn't maintain balance in the system. Any disturbance would cause the Universe to expand or contract, depending on its matter density. In this model, equilibrium doesn't exist, but a force struggle determines whether it expands or con-tracts. This conflict could be eternal or have a past beginning.

George Lemaitre, along with Friedmann, proposed in 1929 that all matter and energy were once concentrated in a singular point, initiating expansion from a massive explosion, known as the Big

Bang Theory (BBT). The BBT gained credibility when Hubble detected a redshift in distant galaxies' light, attributing this effect to high-speed galaxy separation [3].

The Big Bang Theory (BBT) proposes an enigmatic Universe, with challenging-to-determine initial singularity conditions, making a causal explanation of the Universe's evolution difficult. This may have led Thomas Gold and Hermann Bondi to propose the Steady State Theory (SST) in 1948 [4], which models an expanding Universe without a beginning.

As matter expands in the SST, density decreases according to Hubble's law. To counterbalance this loss, Gold proposed a C-field creating one hydrogen atom per cubic meter every $10^{10}$ years throughout the Universe [5]. This ensures an infinite Universe maintains constant structure, isotropy, and homogeneity, preserving the Perfect Cosmological Principle.

The main SST criticism was its lack of energy conservation, similar to the BBT. However, the SST lost credibility with the discovery of cosmic background radiation (CBR), which the BBT explained as a Big Bang remnant. CBR, together with redshift, forms the foundation supporting the BBT.

Some SST authors suggested that the Cosmic Background Radiation (CBR) could result from starlight scattered by interstellar material [6]. However, this explanation faces issues, such as the lack of polarization and the perfect black body characteristics of CBR. Besides this radiation is a perfect black body that could be formed by superposition of radiation with different redshift [7]

The Quasi-Steady State Theory (QSTT, 1993) attempts to bridge the BBT and SST, proposing an eternally expanding and contracting Universe through infinite series of big-bang and big-crunch events [8].

Another alternative comes from Regener, Nernst, Finlay-Freundlich, Max Born, and de Broglie, who considered the Universe is static and infinite, attributing the redshift of starlight to a "tired light" effect [9].

The Quasi-Steady State Theory (QSTT, 1993) attempts to bridge the BBT and SST, proposing an eternally expanding and contracting Universe through infinite series of big-bang and big crunch events [8].

More recent approaches involve describing the observable Universe's structure as a hierarchical distribution of matter using fractal mathematics [10, 11, 12]. and correlation theory, similar to liquid structure analysis [13,14].

These theories showcase the ongoing efforts to understand and describe the Universe, highlighting the evolving nature of cosmological understanding.

## 2. THE MODEL

The distribution of matter in the universe, while appearing non-homogeneous locally, follows a structure with a distribution function inversely proportional to the nth power of the distance from its rotation center at the galaxy and cluster level. This distribution decreases with an exponential factor of ***1.8*** [13, 14, 15]. By extrapolation, it's hypothesized that on a larger scale, it decreases, tending towards a distribution following the inverse square law. The universe may exhibit a fractal structure that adheres to a power law, albeit with scale variance. This proposed distribution model suggests that matter reaches homogeneity at a certain scale. Stars form galaxies, which in turn form clusters, and these clusters group together in a set of clusters, termed superclusters. Each level of these groups rotates around their respective centers of mass, similar to galaxies rotating around their own centers, where matter density could be infinite due to the presence of a super-massive black hole [15, 16].

According to our hypothesis, suggests that the hierarchical levels of matter distribution in the universe reach their maximum at a common rotation center, the Maximum Gravitational Rotation Center (MGRC). This structure, akin to the "bricks" of the universe, could potentially exist infinitely in the infinite universe.

The interaction among these structures is thought to be translational forces, with no higher-level centers of rotation, leading to local dynamics but overall equilibrium. The matter distribution does not follow a radial distribution at this scale, but it tends towards homogeneity.

The density of matter decreases as it moves away radially from the MGRC, with the sepa-ration between stars and galaxies increasing on each scale. s: between stars in the galaxies is of the order of parsecs, and between galaxies in the clusters is in the order of megaparsecs [17], and so successively. The hypothesis aims to align with astronomers' observations about the universe's structure, where homogeneity when the density of the Universe ($9x10^{-27}\ kg/m^3$) is reached, at a certain distance from any MGRC. At this scale, the distribution of matter inside each shell of radius $r \geq R$ remains constant. This is an attempt to align with astronomers' observations about the universe's structure, which suggests that homogeneity is reached at a certain distance from any MGRC, as depicted in the Atlas of the Universe site (see Fig. 1) [18]. This implies that the density of matter becomes consistent across these scales, marking a shift from radial distribution to a more uniform one.

This hypothesis suggests that the universe is a stable infinite system in equilibrium, with multiple MGRCs (Maximum Gravitational Rotation Centers) instead of a dominant one as proposed by the Big Bang Theory. These MGRCs, which could be infinite, are in dynamic equilibrium with each other and the rest of the infinite universe.

This is deduced from the observation that all centers of gravity imply matter rotating around itself, with no dominant MGRC observed. Each MGRC is in dynamic equilibrium with the others, and while gravitational attraction between two centers might dominate, the time it would take for a collision is so long due to the vast distances that collisions are not common.

The stability of this universe is due to the matter of each gravitational center rotating or "falling" to its own center, with a timescale of thousands of years. Although galaxy collisions are possible, they would take a long time. For instance, a predicted collision between Andromeda and the Milky Way is expected in about three billion light-years approximately [19].

This model posits a self-sustaining Universe, where local death of stars or galaxies is balanced by the birth of others in different locations of the same supercluster. This maintains the Universe's homogeneity and isotropy over time. The conservation of matter and energy is ensured through recycling during the Universe's evolution, preserving its state and structure perpetually. The Perfect Cosmological Principle is not assumed, but derived from astronomical observations and physical considerations. In our model, unlike the Big Bang Theory, the Universe, along with its stars and galaxies, has always existed. We hypothesize a constant proportion of matter and radiation in the Universe on average, at any time and place, despite local variations in galaxies or clusters. Given that 0.01% of a star's mass is converted into radiation throughout its luminous life [20], and nearly all Universe's matter is star-made, we consider that 0.01% of the Universe's matter density is radiation.

In this study, we consider an idealized scenario where each star has the same diameter, temperature and luminosity. uniformly distributed within a galaxies with circular disk symmetry (approximating a typical spiral galaxy like the Milky Way. The galaxies interact gravitationally between them forming clusters, superclusters, and so on. We assume a distribution of galaxies inversely proportional to the square of the distance. We use a "stellar dynamics" model, treating each galaxy as a contributor to the overall gravitational field, without needing precise locations, as we did in a previous work [16]. By replacing the distribution of individual galaxies with a smoothed continuum density, we can estimate the gravitational field accurately using Gauss's law. Each galaxy follows a 'collisionless dynamic' around a maximum gravitational rotation center (MGRC), influenced by global gravitational effects and weakly affected by local effects of neighboring stars. We also suggest that any movement in one part of the universe is compensated by the movement of other centers, maintaining a dynamic equilibrium.

## 3. SEMICLASSICAL ANALYSIS

In this work, a semi-classical analysis was employed to derive the gravitational field, considering the significant distances between galaxies and even greater distances between clusters. The gravity force between such structures is relatively small, making this approach suitable. Although General Relativity equations can compute the gravitational force for two particles, their complexity prevents their application to the structures proposed here. The classic analysis, which allows for average calculations of matter distribution using Gauss's law (as done in a previous study [15] for spiral galaxies' rotational velocities), was chosen due to its simplicity and effectiveness for our purposes.

In this analysis, we will divide the investigation into two parts. First, we propose that the redshift radiation from galaxies is a result of gravitational restraint, building on Fritz Zwicki's work, which suggests a gravitational "drag" effect (also known as the tired light effect) [21]. In the second part, we attempt to explain the Cosmic Background Radiation (CBR) as the light from all the stars in an infinite Universe. This light, after traveling through space, undergoes extreme redshift and reaches us without colliding with other stars.

## A) Gravitational Redshift

Our analysis begins by considering the Universe as a continuously distributed matter, which varies inversely proportional to the square of its radial distance from the MGRC

$$\rho = \left(\frac{M}{4\pi R}\right)\frac{1}{r^2} \tag{1}$$

here, $\boldsymbol{M}$ represents the average mass inside a sphere of radius $\boldsymbol{R}$, which marks the minimum length of a shell where matter density transitions to a constant average value. As we move to subsequent layers of similar length (scale), the mass of this distribution increases in a manner proportional to the square of $\boldsymbol{r}$ ($\boldsymbol{r \geq R}$), thereby maintaining a constant density for radii greater than or equal to $\boldsymbol{R}$.

The radial symmetry of this distribution allows, using Gauss's law, to calculate its external gravitational field for $\boldsymbol{r \geq R}$:

$$g = -\left(\frac{GMm}{R}\right)\frac{1}{r}\hat{r} \tag{2}$$

where $\boldsymbol{G}$ is the gravitational constant. Photons traveling through such a gravitational field will have their energy affected, as per the equivalence principle. The frequency of light changes in the presence of a gravitational field: it modifies (diminishes or increases based on its direction relative to gravity).

With the MGRC as our reference point, the distance from the origin to any radiation source (stars in each galaxy) is $\boldsymbol{r'}$. Therefore, photons travel radially from $\boldsymbol{R}$ to $\boldsymbol{r'}$ (from the boundary towards the transition to homogeneity of the matter distribution in the Universe). The calculation of its potential energy will be:

$$\phi = \frac{GMm}{R}\ln\left(\frac{r}{R}\right) \tag{3}$$

where $r = R + r'$, and the 'potential' is positive due to the counteraction of stellar light's direction, which opposes the gravitational field. The hypothetical 'mass' of photons, denoted as $\boldsymbol{m}$, is subsequently annulled.

Indeed, according to Einstein's mass-energy equivalence principle, the total energy ($\boldsymbol{E}$) of a photon, which is a particle of light, is given by the equation,

$$E = mc^2 \quad (4),$$

under the influence of the gravitational potential equation (3), it's modified like

$$E' = E + \frac{GMm}{R}\ln\left(\frac{r}{R}\right) \tag{5}$$

or

$$E' = mc^2 + \frac{GMm}{R}\ln\left(\frac{r}{R}\right) \tag{6}$$

where $\boldsymbol{E}$ represents the initial energy of the photon ($\boldsymbol{h\nu}$), and $\boldsymbol{E'}$ denotes the reduced energy ($\boldsymbol{h\nu'}$). The constant relationship between the energies and frequencies is maintained through Planck's constant ($\boldsymbol{h}$).

Also we know that the redshift is measured with a parameter $z$, defined as

$$z = \frac{\Delta\nu}{\nu} = \frac{(\nu' - \nu)}{\nu} \tag{7}$$

where $\nu$ is the frequency of the light emitted by the source, $\nu'$ is the light with redshift received. We can rewrite this expression as

$$1 + z = \frac{\nu'}{\nu} \tag{8}$$

multiplying by the Planck's constant $h$, numerator and denominator, we can express this relation as

$$1 + z = \frac{E'}{E} \tag{9}$$

Then, replacing the value of $E$ and $E'$ of the previous equations (4) and (6) respectively, in eq. (9)

$$1 + z = \frac{mc^2 + \frac{GMm}{R}\ln\left(\frac{r}{R}\right)}{mc^2} \tag{10}$$

therefore

$$1 + z = 1 + \frac{GM}{c^2R}\ln\left(\frac{r}{R}\right) \tag{11}$$

and we obtain

$$z = \frac{GM}{c^2R}\ln\left(\frac{r}{R}\right) \tag{12}$$

The gravitational redshift expression we've obtained will be considered a general form of Hubble's law, which we will try to prove next. This logarithmic expression can be developed using Taylor series

$$z = \frac{GM}{c^2R}\left[\frac{r}{R} - \frac{1}{2}\left(\frac{r}{R}\right)^2 + \frac{1}{3}\left(\frac{r}{R}\right)^3 - \cdots\right] \tag{13}$$

The linear expression serves as the initial approach for the potential values near its periphery $\boldsymbol{R}$, where the matter density reaches or approximates a constant value. Consequently, in this section, we will adopt a linear approach.

$$z = \frac{GM}{c^2R^2}r, \tag{14}$$

In this approach, we aim to measure the gravitational effect on light at a large scale, where the Universe is homogeneous. We can extrapolate this expression for distances greater than $\boldsymbol{R}$, assuming that gravity's effect remains constant at these scales. This global effect of gravity could theoretically 'tires' the light uniformly, regardless of direction or distance, due to the isotropic nature of the Universe.

From the Hubble law, we know that

$$cz = Hr \tag{15}$$

and therefore

$$H = \frac{GM}{cR^2} \tag{16}$$

It's noteworthy that the last equation could potentially explain the deviations in the predicted trajectories and velocities of Pioneer 10 and 11, a phenomenon known as the Pioneer anomaly. Researchers studying this effect discovered a constant sunward acceleration of ***(8.74 ± 1.33)*** × ***10*** $^{-10}$ ***m/s***$^2$ [21]. Interestingly, the magnitude of this quantity is roughly equivalent to the product of the speed of light and the Hubble constant ***(Hc=7.23*** × ***10*** $^{-10}$ ***m/s***$^2$ ). Therefore, we could rewrite Eq. (16) as

$$Hc = \frac{GM}{R^2} \tag{17}$$

The right side of this equation represents the deceleration that any body with mass, or even electromagnetic radiation like the light, could experience due to the gravity of matter distribution, according to our model at a radius ***R***. This is commonly known as a "drag" effect [22]. Therefore, this equation explains why the product of the Hubble constant ***H*** and the speed of light ***c*** gives the magnitude and its physical unit of this deceleration.

Now, the value of ***R*** can be determined using equation (17). The necessary data includes ***M*** and ***H***, which can be used to calculate the first definition of density. Assuming that on this scale our density function (eq. (1)) should correspond with the average density of the Universe, as we are in the transition zone to homogeneity, we can calculate ***M*** by multiplying the average density ($\rho_0$) by the volume of the radio ***R*** occupies.

$$M = \left(\frac{4\pi R^3}{3}\right)\rho \tag{18}$$

Now, if we substitute equation (18) into equation (17)

$$H = \frac{4\pi G R \rho}{3c} \tag{19}$$

and clearing ***R***, we obtain

$$R = \frac{3cH}{4\pi G R \rho} \tag{20}$$

Now, we could obtain the Hubble constant (***H***) directly from the graph of Redshift versus Luminosity Distance diagram [23] (in the scale of Gigaparsecs) using a linear approximation. This gives us an approximate observational value of ***H = 73.8 km/s/Mpc***. Substituting these values in equation (20), we obtain ***R=3x10*** $^{26}$ ***m,*** which is the radius of the MGRC.

Below, we can derive an expression for the gravitational redshift using the known data. By substituting the value of ***H*** from equation (19) into equation (15), we obtain:

$$z = \frac{4\pi G R \rho}{3c^2} \tag{21}$$

and substituting *z* value in Eq. (11)

$$1 + \frac{4\pi G R \rho}{3c^2} r = \frac{\nu}{\nu'} \tag{22}$$

we clear $\nu'$ to obtain

$$\nu = \left(1 + \frac{4\pi G R \rho}{3c^2} r\right) \nu' \quad (23)$$

and finally, we find a formula for the gravitational redshift of the frequency of the photons.

## B) Cosmic Background Radiation

According to our hypothesis, which we will attempt to prove, the Cosmic Background Radiation (CBR) is caused by the spectral radiation of all the stars, grouped in galaxies and in clusters of galaxies, across our infinite universe.

The calculations we're about to perform involve significant idealization. We're using a model of the Universe where galaxies are homogeneously distributed, ignoring their clustering into clusters or superclusters. We assume each galaxy has circular disk symmetry, with a diameter similar to the Milky Way galaxy (because it is an average galaxy), which is the galaxy that contains our solar system. The Milky Way galaxy is estimated to have a mass of about 1.5 trillion (***$1.5x10^{12}$***) times the mass of our sun [24]. Since the mass of our sun is approximately ***$2 \times 10^{30}$ kg***, doing the math gives us a total mass for the Milky Way of around ***$3 \times 10^{42}$ kg***. This is a rough estimate, as the exact mass is difficult to measure due to the vast size and unknown contents of the galaxy.

But not all stars are similar to our sun (although we are going to start the calculation with this one), we'll consider its distribution percentages according to the Harvard Spectral Classification: stellar radiation characteristics based on star types, their temperatures and mass. Taking all these factors into account, despite an extreme idealization, we believe that we can achieve an accurate approach that aids in proving our hypothesis. This is because it indeed possesses these charac-teristics on a large scale, applicable to the universe, where we will conduct our calcu-lations.

Photons from these stars traverse freely through space until they reach us, unless they collide with a galaxy in their path and are absorbed by another star or interstellar material. This trajectory is commonly referred to as a 'mean free path' in the kinetic theory of ideal gases. Here, photons are considered analogous to particles that navigate obstacles -static, uniformly distributed galaxies throughout the universe. The bottom limit of this trajectory is termed as such [25]. This quantity can be computed using the following expression:

$$bottom - lim = \frac{Vol-occupated-for-galaxy}{Transversal-Section-Area} \quad (24)$$

1.

This limit can be approximated, given our understanding of the average density of matter in our universe. With this, we can calculate the average volume occupied by each galaxy in the universe (***$M_{gal}$***). The average density of the Universe can be defined as the volume of space that a galaxy (***$V_{ocup}$***) with average mass would occupy, if these galaxies were uniformly distributed in the space:

$$\rho = \frac{M_{gal}}{V_{ocup}} \quad (25)$$

Substituting the average density, ***$9x10^{-27}$ $kg/m^3$***, and the average galaxy mass (approximately ***$3x10^{42}$ kg***), we have:

$$9 \times 10^{-27} = \frac{3\times10^{42}}{V_{ocup}} \quad (26)$$

and solving this equation we obtain

$$V_{ocup} = 3.33 \times 10^{68} m^3.$$

We don't know the average cross-sectional area of the galaxies, but we do know that most galaxies are 1 to 100 kiloparsecs (kpc) in diameter [26]. Therefore, for the purpose of this analysis, we need to make some assumptions. First, let's consider a minimum diameter for this analysis. We'll assume that our galaxy has spherical symmetry, with a diameter equal to ***1 kpc***, and can be calculated as follows: ***Cross-Sectional-Area = π*(diameter/2)***$^2$. Then, the cross-sectional area of a galaxy is approximately ***7.48 x*** $10^{38}$ $m^2$. Now, substituting this in equation (24), and now we can to obtain the lower limit: $l=4.41x10^{30}$ ***m***. This value represents the bottom limit, indicating the maximum distance that the light from a star in a given galaxy can travel to another galaxy. In other words, this magnitude signifies the distance spanned by a given galaxy, encompassed by all other galaxies. This implies that if you possess a highly powerful telescope to reach this limit and focus in any direction, you would invariably detect a star. Consequently, no ray of light from any other star beyond this distance could traverse this limit. The path of this ray of light has been historically referred to by astronomers as a line of sight [25]. The average decay of photons absorbed by other stars during their journey from a distance r is provided by ***exp(-r/l)*** [25]. We're not taking into account the space between the stars here, as the presence of gas and dust clouds within galaxies often conceals this emptiness. This matter, which makes up approximately ***20%*** of the Milky way galaxy, is also included in our consideration [27]

The total radiation of the universe, or at least that which reaches us up to the bottom limit, can be calculated on average by dividing space into infinitesimal spherical layers, with the origin being any center of a MGRC. The radiation intensity, or the number of photons, will be directly proportional to the number of stars contained within each infinitesimal layer. The volume of such a layer can be mathematically expressed as

$$4\pi r^2 dr \tag{27}$$

The number of galaxies contained within each infinitesimal volume is determined by dividing that volume with respect to the average volume occupied by each galaxy in the universe

$$\frac{4\pi r^2 dr}{1.19\times 10^{68}}. \tag{28}$$

This quantity will provide the average number of galaxies within each infinitesimal layer, expressed as a quadratic function of its distance to a MGRC. Given that each galaxy contains approximately $1.5x10^{12}$ stars, the total number of stars within each layer is obtained by multiplying the aforementioned expression by this number

$$\frac{1.5\times 10^{12}}{8.341\times 10^{68}} 4\pi^2 dr \tag{29}$$

Besides, each star emits solar spectral radiation, and the expression for the number of photons per unit area, unit frequency, and unit time is

$$\frac{2\nu^2}{c^2}\frac{1}{e^{\frac{h\nu}{kT}}-1} \tag{30}$$

the function, where *k* represents Boltzmann's constant and c denotes the speed of light, is intended to be multiplied by the star's surface area. In our model, the star's surface area is analogous to that of our sun, approximately $4\pi R^2$ (where ***R*** is the radius of the sun, then $6x10^{18}$ $m^2$) This results in the total number of photons emitted by the star per second for each frequency in its spectrum

$$\frac{2\nu^2}{c^2}\frac{6\times10^{18}}{e^{\frac{h\nu}{kT}}-1} \tag{31}$$

Now, this amount is due to multiply by the total number of stars in each shell, calculated previously in equation (29), to obtain a spectral radiation of photons as a function of distance ***r***

$$\frac{2\nu^2}{c^2}\frac{6\times10^{18}}{e^{\frac{h\nu}{kT}}-1}1.8\times10^{-57}\times4\pi^2 dr \tag{32}$$

Indeed, as each star in each layer emits radiation from a distance r of MGRC, its light spreads over a spherical surface area with this radius. Therefore, the previous expression should be divided by $4\pi r^2$ to account for this

$$\frac{2\nu^2}{c^2}\frac{1.08\times10^{-38}}{e^{\frac{h\nu}{kT}}-1}dr \tag{33}$$

This effect is compensated by the volume of each layer, meaning its contribution remains a constant quantity for each shell, independent of the distance ***r***.

Furthermore, the quantity of photons arriving from space to us is reduced due to absorption by stellar objects, as explained earlier. This average reduction follows an exponential pattern, and we have

$$dN(r,\nu)=\frac{2\nu^2}{c^2}\frac{1.08\times10^{-38}}{e^{\frac{h\nu}{kT}}-1}e^{-\frac{r}{l}}dr \tag{34}$$

Besides, the radiation intensity received from each layer to a distance of ***r*** is obtained by multiplying the previous expression by the energy of each photon, corresponding to its frequency. The energy of each photon is modified along its route due to gravitational redshift, as per equation (23).

$$dI(r,\nu)=dN(r,\nu)h\nu' \tag{35}$$

Then, the spectral radiation for each shell is written as

$$dI(r,\nu)=\frac{2h\nu^2\nu'}{c^2}\frac{1.08\times10^{-38}}{e^{\frac{h\nu}{kT}}-1}e^{-\frac{r}{l}}dr \qquad \text{witn} \qquad \nu=\left(1+\frac{4\pi R\rho G}{3c^2}\right)\nu' \tag{36}$$

Therefore, for any distance ***r***, the spectral radiation received from the stars in the Universe at this distance can be expressed as

$$\sigma(r,\nu')=\frac{2h\nu'^3}{c^2}\left(1+\frac{4\pi R\rho G}{3c^2}r\right)^2\frac{1.08\times10^{-38}}{e^{\frac{h\nu'}{kT}\left(1+\frac{4\pi R\rho G}{3c^2}r\right)}-1}e^{-\frac{r}{l}} \tag{37}$$

This is an energy density of a spectral radiation density that arrives linearly redshifted to us, as shown in the expression for ***ν*** (Eq. (23)). We assume that this radiation reaching us, ***ν'***, is the spectral radiation of the CBR.

Now, we are interested in obtaining a graph of the energy density as a function of ***r***. We can rewrite this equation (37) as

$$\sigma(r,\nu')=1.59\times10^{-88}\frac{(1+\propto r)^2\nu'^3}{e^{\beta(1+\propto r)\nu'}-1}e^{-\lambda r} \tag{38}$$

where $\propto=\frac{4\pi GR\rho}{3c^2}$, $\beta=\frac{h}{kT}$ and $\lambda=\frac{1}{l}$.

Now, all these parameters can be calculated and should be scalable because the values of ⟨ and ® should not be too small, as computer calculations have insufficient precision. Considering that the distance ***r*** on a scale of ***$10^5$*** megaparsecs (written as ***$r=3.086x10^{27}r'$***). Besides, a star of type ***G***, like our sun, and substituting the constant values, we obtain numerical values for ***$\alpha=25.84$***, ***$\beta=8.42x\ 10^{-15}$*** and ***$\lambda=0.0006984$***.

We will fix ν' at a value corresponding to the frequency (***$\nu'=1.5x10^{11}$***) where the Cosmic Background Radiation (CBR) has maximum intensity. Upon substituting these values into equation (38), we finally have

$$\sigma(r,\nu') = 5.36\times 10^{-55}\,\frac{(1+\propto r)^2}{e^{\delta(1+\propto r)}-1}\,e^{-\lambda r}$$

where **$\delta=\beta\nu'=1.26x10^{-3}$**. With this consideration, the function now only depends on the variable ***$r'$***. We can plot its graph in terms of ***$r'$***, which shows all possible ***$\nu$*** values that contribute (for all stars at any distant ***$r'$***) with a fixed ***$\nu'$*** (refer to Figure 3). Although here we consider a star similar to our sun, the energy density distribution is similar for any other star type in the Hertzsprung-Russell diagram (HRD).

In this graph, we observe that the intensity peak is approximately similar to the bottom limit's distance. This peak also aligns with the frequency of maximum solar radiation (***$3.5x\ 10^{14}$ Hz***). The redshift (***$\Delta\nu$ =1,570 Hz***) for this distance is approximately the same as the quantity required to achieve the CBR frequency. We consider this region as the visible Universe's limit, as its frequency, with extreme redshift, falls outside the visible spectrum (supposedly from here, the radiation reaches us as CBR).

Therefore, we must calculate the CBR from this limit to infinite. Here, we consider infinite as the distance where radiation contribution is nullified by the negative exponential of the bottom limit. For numerical calculation, this approximation is valid. Similar to energy density, we set a numerical ν' value and integrate this expression. We calculate it for various ***$\nu'$*** values, ranging from ***$1x10^9$ Hz*** to ***$1x10^{12}$ Hz***. With this, we add the contribution for ***$\nu'$***, from any ***$\nu$*** of the spectrum of all stars redshifted, and at a specific distance beyond the visible Universe.

To solve the integral of equation (38) analytically, we need to make the following approximation, $\beta(1+\alpha r)\nu' \gg 1$, therefore

$$\sigma(r,\nu`) = 1.59x10^{-88}\nu'^3(1+\alpha r)^2 e^{-\lambda r-\beta(1+\alpha r)\nu'}$$

We must consider that every shell contributes a different quantity of photons, ***$\nu'$***, to the CBR spectral radiance for the same frequency, ***$\nu$***, due to their gravitational redshift being a function of distance. The intensity (or number of photons) contribution is maximum near the bottom limit, where radiation coincides with the frequency of the maximum intensity of the solar spectral radiation. However, this bottom limit also depends on the frequency of the maximum spectral radiation for each type of star in the HSC, as the frequency ***$\nu'$*** depends on the distance traveled by its light. Therefore, to integrate all the contribution an approximated calculation should be performed, considering that the distance ***r*** on a scale of ***$10^5$ mega parsec***, written as ***$r=3.086x10^{27}r'$,***

$$I(r`,\nu`) = 4.9x10^{-61}\nu'^3\times\int_{ri}^{rf}(1+\alpha r)^2\,e^{-\lambda r-\beta(1+\alpha r)\nu'}dr'$$

where ***ri = 30*** to ***rf = $5*10^5$*** scale units, considering the Spectral radiation density of the Universe (Figure 3). For radii smaller than ***ri***, their contribution is not considered, as this radiation is perceived as the spectrum of each star with a specific location, not as background radiation. The ***rf*** radius is up to infinity, although the bottom limit is obtained at the scale of ***$5*10^5$***.

By expanding the binomial into a square and integrating each part in terms of ***r***, while treating ***v*** as a constant, we obtain the CBR intensity as a function of $\boldsymbol{v'}$ :

$$I(\nu') = 4.44 \times 10^{-61} \times \nu'^3 e^{-\beta\nu'} \left[\frac{e^{ar}}{a}\left\{1 + \propto^2 \left(r^2 - \frac{2r}{a} + \frac{2}{a^2}\right) + 2 \propto \left(r - \frac{1}{a}\right)\right\}\right]_{ri}^{rf}$$

where $a = -\lambda - \propto \beta\nu'$.

When calculating the contribution for to calculate the average mass of each star type, you need to multiply the percentage factor of its distribution by its average mass. Here are the correct values [28]:

- M-type stars: (0.76) x (0.45 M_⊙)
- K-type stars: (0.12) x (0.6 M_⊙)
- G-type stars (like the Sun): (0.076) x (1 M_⊙)
- F-type stars: (0.03) x (1.2 M_⊙)
- A-type stars: (0.006) x (1.75 M_⊙)
- B-type stars: (0.0013) x (9 M_⊙)

Note that we are not considering O-type stars, as their contribution is negligible due to their small fraction and frequency. To find the average mass, simply multiply the distribution percentage by the average mass for each star type and add them together. Finally, we add each contribution to obtain a graph shown in Fig. 4.

Additionally, the average temperature for this spectral radiation intensity can be calculated using the Stefan-Boltzmann law for radiation, as stated in equation [29]

$$\rho c^2 = \frac{4\sigma}{c} T^4 \tag{39}$$

where $\boldsymbol{\rho}$ represents the density of the radiation in the universe, which is, on average, 0.01% of all matter in our universe, or $\boldsymbol{1.67x10^{-31}\ kg/m^3}$. Here, ***c*** is the speed of light, $\boldsymbol{3x10^8\ ms^{-1}}$; and $\boldsymbol{\sigma}$ is the Stefan-Boltzmann constant, $\boldsymbol{5.67x10^{-8}\ Jm^{-2}K^{-4}s^{-1}}$. With all this data, we find that the temperature is $\boldsymbol{2.11^0\ K}$, which is quite close to the $\boldsymbol{2.73^0K}$ of the ***CBR*** obtained from observational measurements.

## Curvature of the Hubble Diagram

Even though we do not believe in the expansion of the Universe, we must still explain the curvature of the redshift obtained from the measurement of the luminosity of distant Supernovae [23], which has been interpreted as the Universe accelerating (or decelerating?). In this section, we utilize the approximated formula of the Hubble's law, as shown in equation (15):

$$cz = Hr$$

The Hubble's law is valid for distances within or near the periphery of ***R***. Beyond this limit, a linear approximation must be used, as the Universe is homogeneous and isotropic on a larger scale, meaning the effect is the same in any direction or length. Using this function, we obtain the graph shown in Fig. 5, which demonstrates a relatively good fit to the data, as the curve falls within the error bars to within an order of magnitude [30].

## 4. DISCUSSION

We have demonstrated that the redshift of light can be explained as a gravitational effect of constraint, which depends on the form of the gravitational potential caused by the distribution of matter in the Universe. Our analysis began with a hypothesis, based on observational data and physical considerations, that the structure of the Universe has a matter distribution inversely proportional to the square of the distance to the gravitational center (MGRC). This distribution is in dynamic balance with an infinite number of similar distributions. This hypothesis is supported by the following deductions: we think that, at the scale of clusters or superclusters, this is the largest unit of the Universe in which matter is grouped, akin to the bricks of the universe. The interaction between these units occurs through translational force, with a dynamic balance in place as there is no dominant gravitational center or a center of rotation for other MGRCs on a larger scale. The structure could resemble a fractal, with a variance in scale of its matter distribution based on the distance inversely proportional to some potential, having a radius limit $\boldsymbol{R}$. As the potential approaches the second order, the Universe moves towards an average homogeneous state, despite the seemingly random distribution on smaller scales.

Performing a 'stellar dynamics' analysis on a smoothed distribution of matter allows for the calculation of its average gravity at the R scale. We conducted an analysis using the Equivalence Principle, the Energy Conservation Principle, and Einstein's Mass-Energy relation, which led us to determine its gravitational redshift, $z$. From this, we derived a mathematical expression for the $\boldsymbol{H}$ constant and a generalization of Hubble's law from first principles. This generalized Hubble's law enables us to explain the curvature of the Hubble diagram without requiring the hypothesis of the accelerating Universe, as per the Big Bang Theory (BBT) [31, 32].

The value of $\boldsymbol{H}$, obtained from the linear approximation of the curvature in the Hubble diagram through astronomical observations at the Gigaparsec scale, and the known mass density of the Universe, is used to calculate the numerical value of $\boldsymbol{R}$. This value aligns roughly with the distance or scale where the Universe becomes homogeneous. Upon introducing this $\boldsymbol{R}$ value into eq. (17), we approximate the deceleration of the Pioneer anomaly. These calculations lead to a linear expression dependent on distance $\boldsymbol{r}$ for determining the gravitational redshift of light.

We have determined the distribution of stellar radiation in a hypothetical, boundless universe. Our model is highly idealized, assuming stars are like the sun and uniformly distributed in spherical galaxies of consistent diameter throughout all space. We disregard clusters of galaxies or superclusters. We also assume the radiation reaching us from the entire universe is constant in intensity at all times and spaces, suggesting an infinite, eternal, and self-sustaining universe. After considering gravitational redshift and the mean free path of this radiation, we calculated the total average radiation at any universal location and time. The results indicate a radiation distribution in frequency and intensity similar to the Cosmic Background Radiation (CBR). Using the Stephan-Boltzmann law for radiation, we computed the average universe temperature, finding a temperature just $0.6^{0}$ K less than observational data. This is a notable result, considering the approximations in our data. It suggests that the CBR might be due to radiation from stars in an infinite universe. This precision was not achieved by the Big Bang Theory, which predicted a temperature of around $50^{0}$ K before the CBR's discovery.

Our findings also address the Cheseaux-Olbers paradox, acknowledging Cheseaux's historical contribution (see [25]). The bottom limit we propose prevents the universe from being inundated with radiation, as it blocks photons from the infinite universe from reaching us. Additionally, the reduction in radiation intensity is due to the significant energy loss of these photons caused by extreme gravitational redshift. Interestingly, the CBR could potentially be the radiation predicted by this paradox.

Our calculations have been highly idealized, allowing us to make approaches that facilitate calculations. Despite this, our results are a good approximation. This could be because our universe model might be correct, and despite not considering matter distribution randomness, it approximates the average value observed in the universe at large scales. We've approximated data, such as the universe's density or average galaxy radius, making conclusive results challenging to obtain. However, if our analysis lacked validity, the astronomical figures would have led to an absurd result. Our result is approximately two orders of magnitude from the CBR's intensity and has a similar frequency.

Additionally, besides the CBR and redshift, another significant evidence supporting the Big Bang Theory (BBT) is the observed asymptotic decrease to zero of heavy element abundance in older stellar objects, as well as a constant 24% abundance of ***He*** in total mass [33, 34]. According to our model, this can be explained by the universe's self-sustainability. Most galaxies' stars are of the Asymptotic Giant Branch type, evolving into Red Giants. In their final stage, these stars eject outer layers, forming planetary nebulae, while their cores evolve into White Dwarfs. This ejected material, rich in heavy elements and high in ***H*** and ***He***, is similar to the stars' photospheric composition (75% ***H***, 25% ***He***). Its composition remains relatively unchanged during evolution since nuclear reactions occur in the core and surrounding layers. Consequently, this ejected material is recycled, contributing to the formation of new stars with these heavy elements in their structure. This process aligns with the self-sustainability concept in the universe.

As per the Big Bang Theory (BBT), the first stars formed after the Big Bang, known as Population III stars, wouldn't contain any heavy elements, as they didn't exist at the universe's origin. Direct evidence of their existence is still lacking, though some researchers claim to have found traces in a gravitationally lensed galaxy [35]. Subsequently, stars are believed to have become more metal-enriched, as heavy elements from planetary nebulae of previous generations were recycled during new star formations (Population II stars). Over time, stars evolved, resulting in present-day stars with high metallic content (Population I stars). Despite this, stars with low heavy element quantities have been observed in the Milky Way's spiral arm, with greater amounts found in the bulge and halo.

The gradient in metallicity has been attributed to a higher concentration of stars in the galaxy's center, resulting in more heavy elements being recycled into new generations. However, this argument is questionable since star formation is more frequent in this region. Observations from the Chandra X-ray Observatory suggest that the supermassive black hole in the Milky Way galaxy's center aids in star formation [36], and research led by Yair Krongold et al. indicates that strong winds from this black hole expel heavy elements [37]. Thus, this argument does not seem to support the BBT.

Another claimed test for the BBT is the time dilation phenomenon in supernova light curves [358, 39]. However, our model can also explain this effect perfectly, attributing it to gravity's impact on time dilation, aligning with the equivalence principle.

In conclusion, we have presented in this paper an alternative explanation to the redshift and the CBR, offering a third option to the two already known: the Big Bang Theory and Steady State Theory. We propose a new structure and dynamics for a Static Universe: infinite, eternal, and self-sustainable. We believe our model is simpler and more coherent than existing ones, and falsifiable in the Popperian sense of the term.

In line with our hypothesis, the Cosmic Background Radiation (CBR) should contain information about the universe's structure. Consequently, we plan to perform a computational calculation considering all scales of the proposed universe's structure, including its randomness and gravitational redshift effect. If our hypothesis is correct, we should be able to calculate the CBR's anisotropy. This could potentially serve as a method to refute our universe model.

We acknowledge that our model is a rough approximation. Our initial aim follows John Wheeler's First Moral Principle: never perform calculations until you know the answer. Hence, we first conduct a semi-quantitative analysis, and in the future, we plan to perform a more precise mathematical and computational analysis. However, we also need a more accurate model and better observational data. We must be mindful of the limiting factors, both human and technical, as well as epistemological, as demonstrated by Chaos Theory. We should be cautious of theories that offer models with high precision, as with sufficient math and parameters, any curve can be adjusted. As Einstein said, our model should have the smallest possible number of arbitrary parameters. This is the issue in contemporary Science, as expressed by Abraham Moles [40]: we live in the era of 'Quantofreny', an obsession with the precision of measured quantities, under the belief that our models can achieve any precision.

We do not claim that our work provides a superior explanation of the Universe compared to other models. Instead, it aims to present another potential explanation about the Universe. We conclude that, when it comes to anything, but particularly the Universe, absolute certainty is hard to come by. Hence, we have opened up the spectrum of all possible explanations.

## ACKNOWLEGMENTS

E.L.S. is very grateful to Josè Abdala Helayel Neto, Andre Koch Torres Assis, Martín López-Correidora, Bruce Evry and Daniel Acosta Avalos for their very careful reading and helpful discussion of the manuscript. E.L.S would like to acknowledge the partial support of Brazilian agencies: Conselho Nacional de Desenvolvimento Científico e Tecnológico (CNPq), and Centro Latino Americano de Física (CLAF).

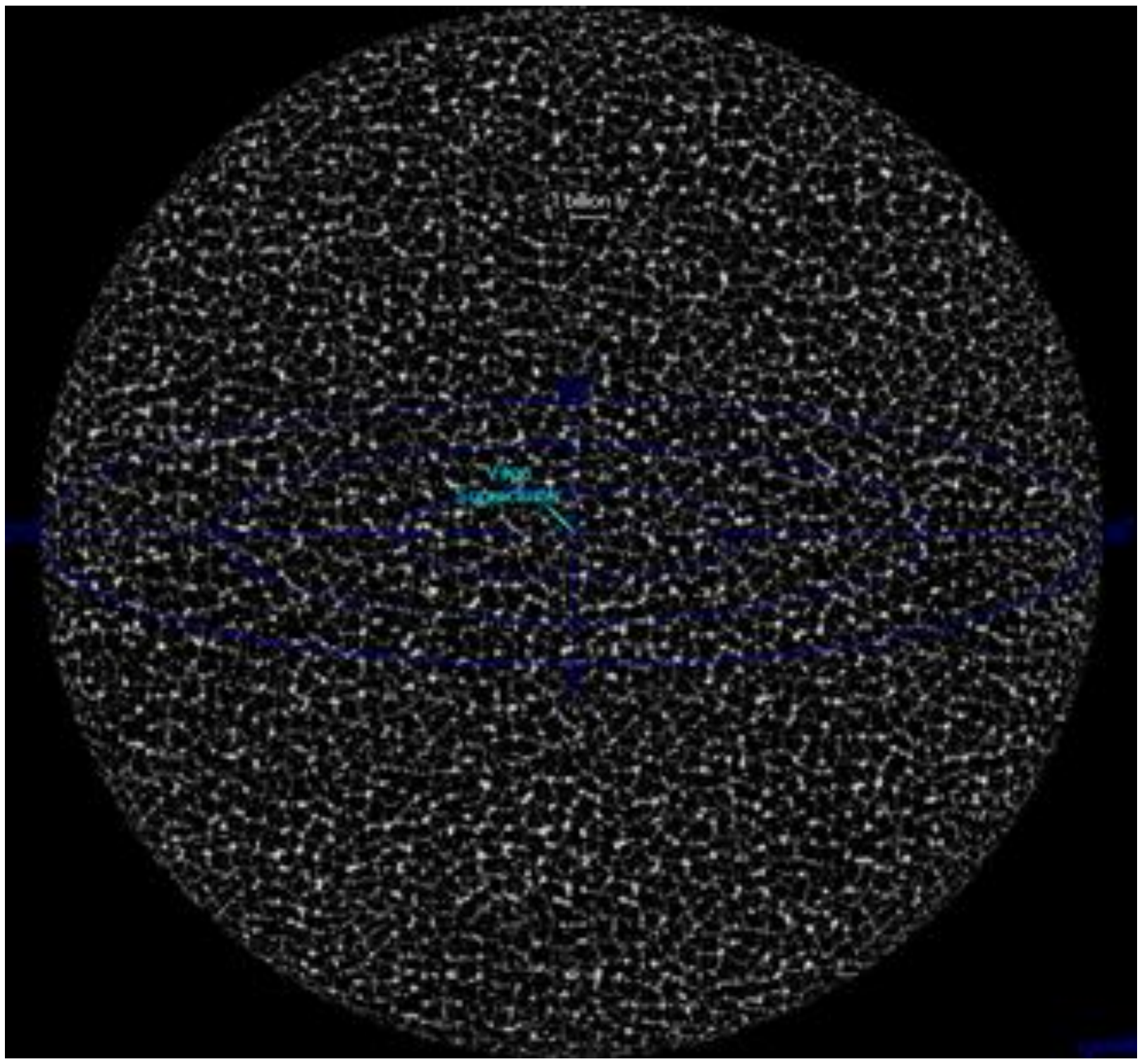

Figure 1. Atlas of the visible Universe (14 billions light years of the sun). To this scale the Universe is fairly uniform (Atlas of the Universe, from Richard Powell).

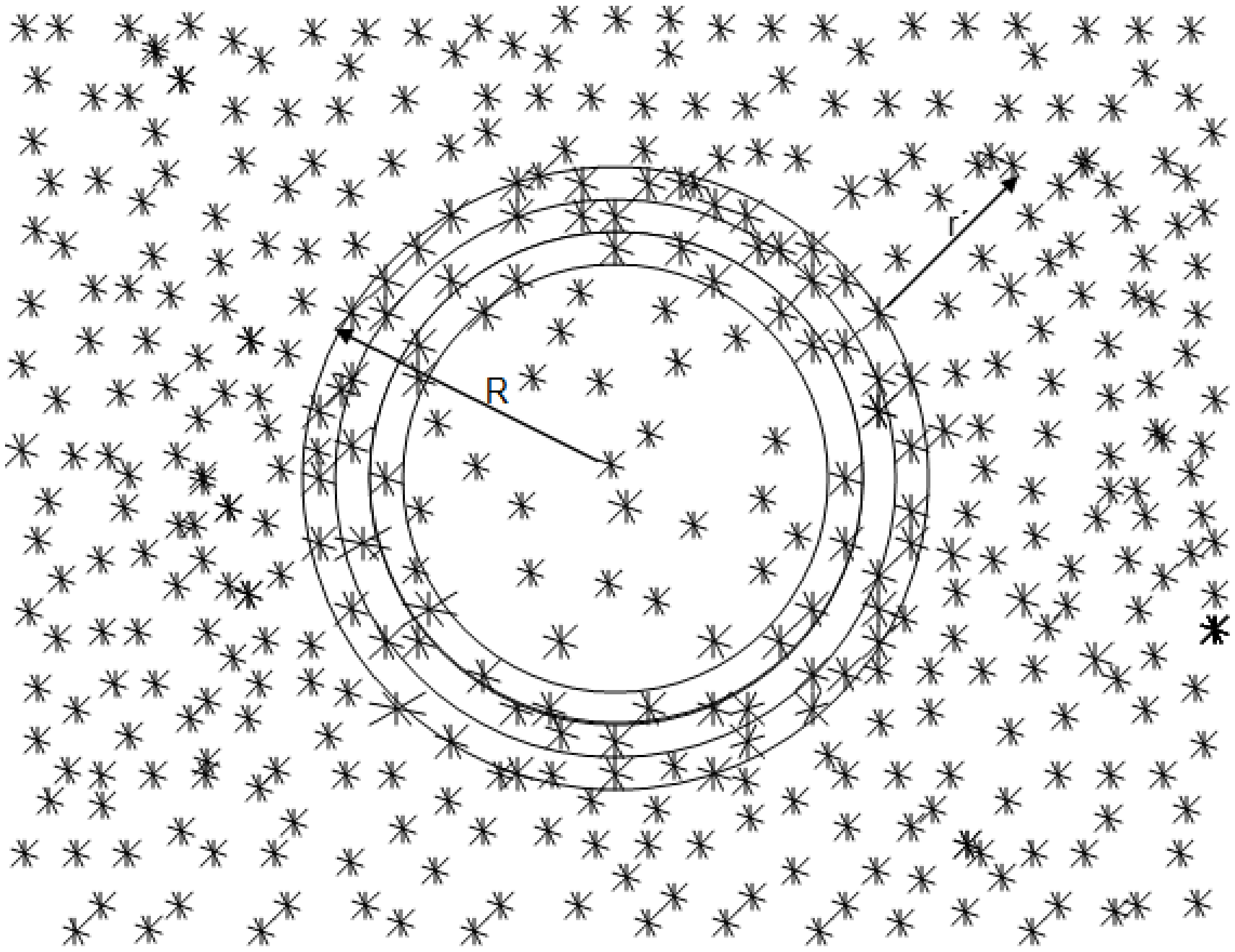

Figure 2. Schematic diagram showing how the galaxies are distributed in the Universe until to reach a homogeneous distribution to the distance *R* from our galaxy.

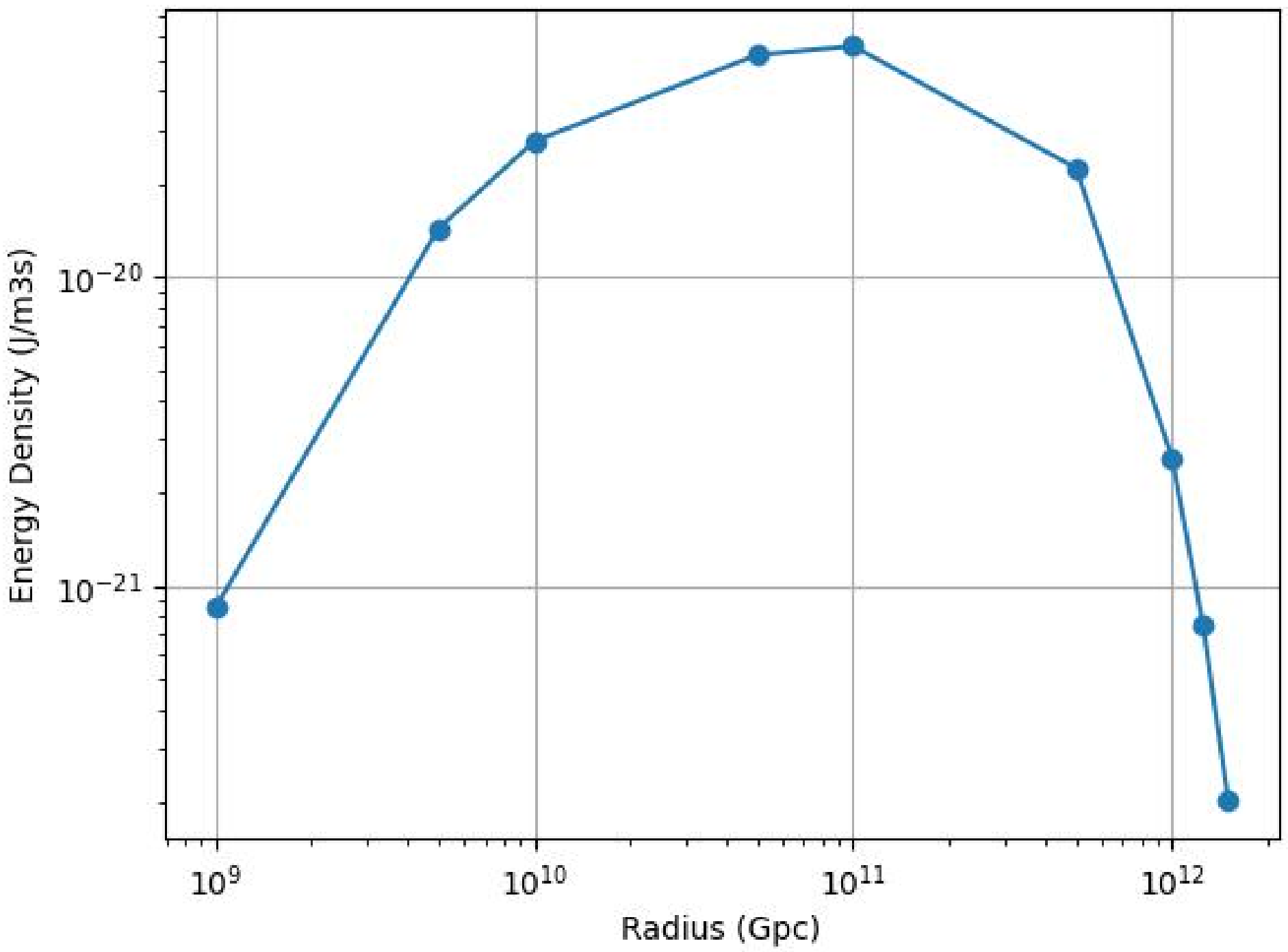

Figure 3. Spectral radiation density of the Universe in function of the *r* distance from the MGRC (Maximum Gravitational Rotation Center), calculated according to our model.

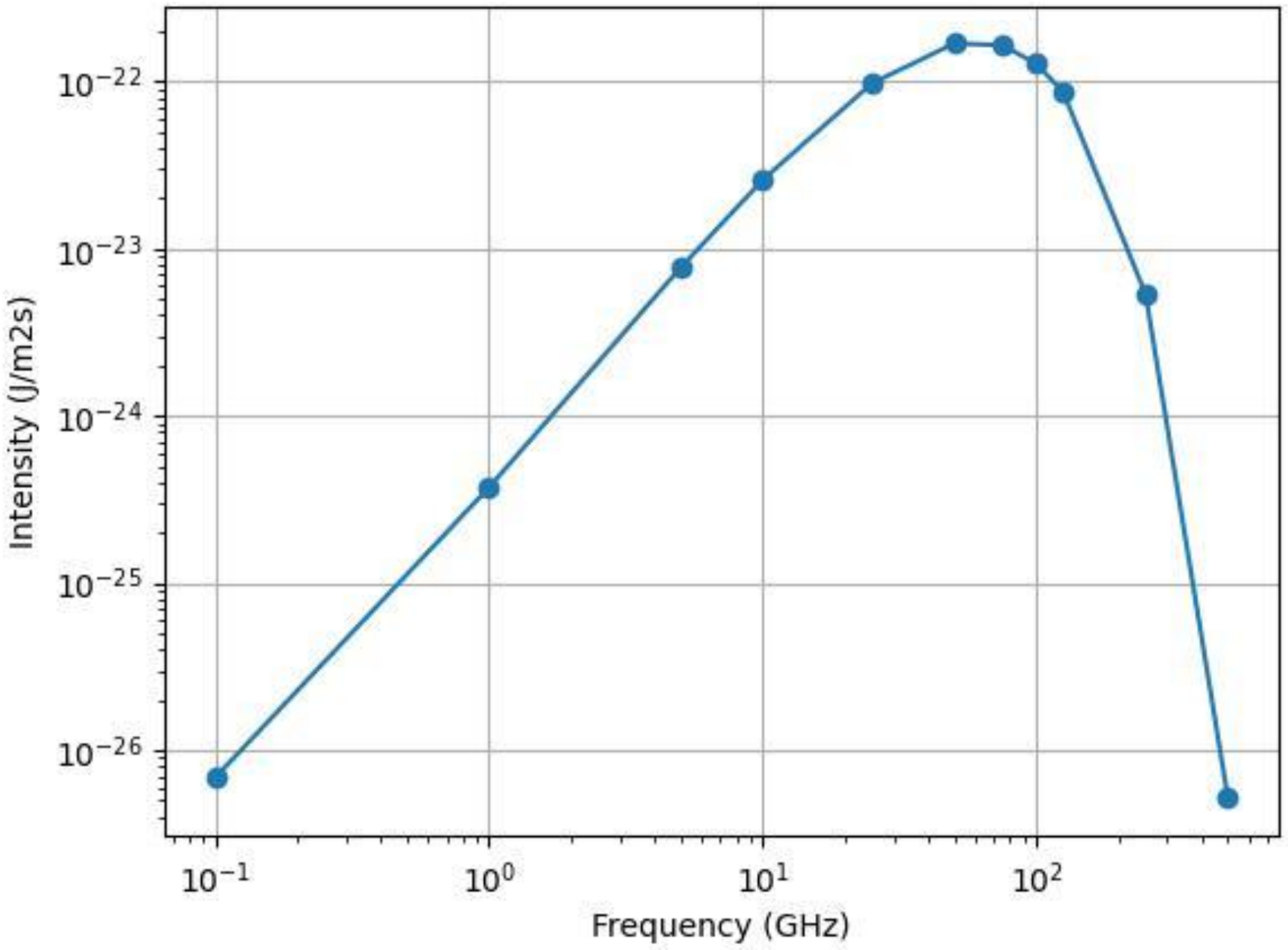

Figure 4. Spectral radiation intensity of the Universe, calculated according to our model and that approximates to the CBR obtained from astronomical measurements.

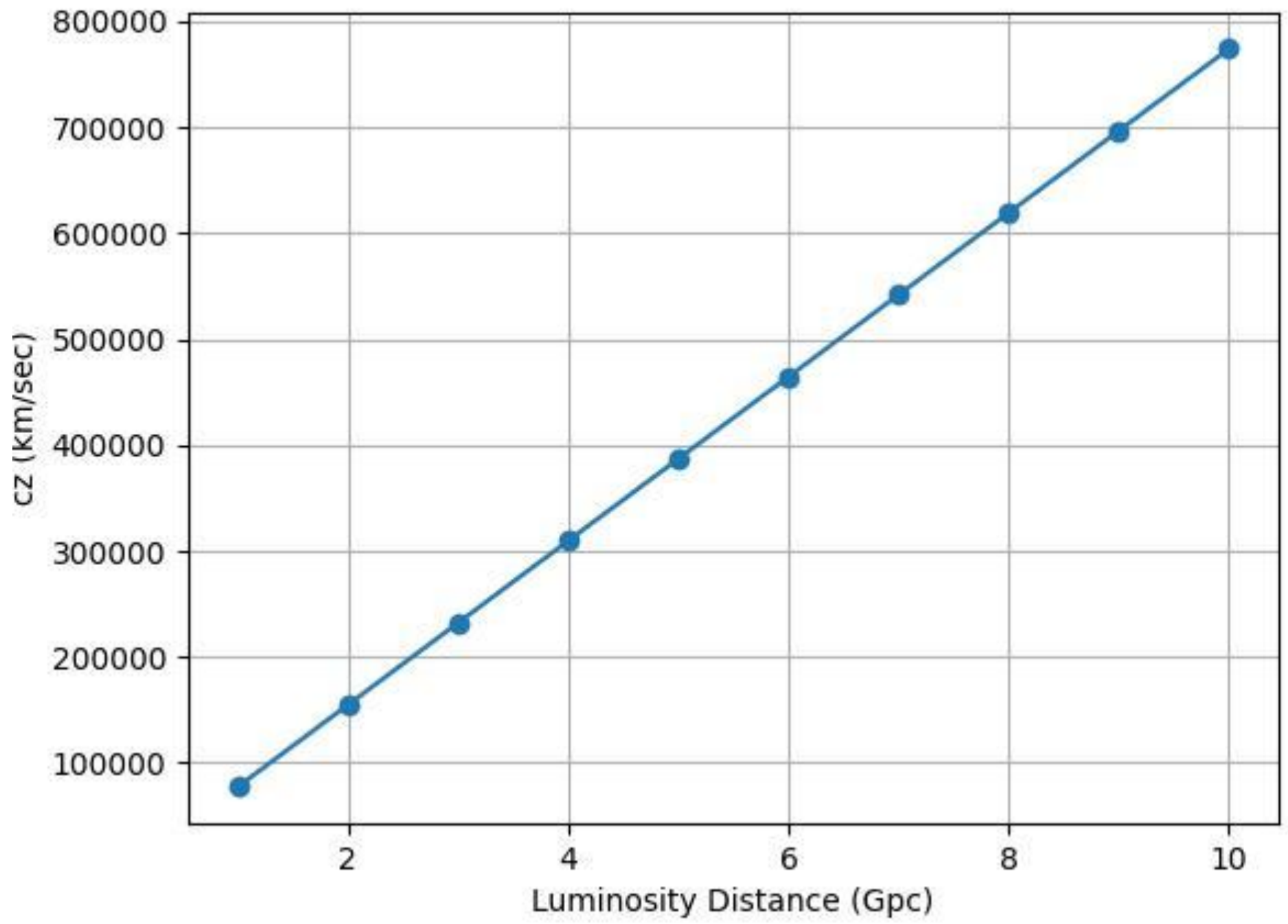

Figure 5. Redshift of the distant Supernovas vs. luminosity distance.